\documentstyle[12pt]{article}

\newcommand{\resection}[1]{\setcounter{equation}{0}\section{#1}}

\newfont{\twelvemsb}{msbm10 scaled\magstep1}
\newfont{\eightmsb}{msbm8}
\newfam\msbfam
\textfont\msbfam=\twelvemsb
\scriptfont\msbfam=\eightmsb
\catcode`\@=11
\def\Bbb{\ifmmode\let\next\Bbb@\else
  \def\next{\errmessage{Use \string\Bbb\space only in math mode}}\fi\next}
\def\Bbb@#1{{\fam\msbfam{{#1}}}}

\newtheorem{theorem}{Theorem}[section]
\newtheorem{lemma}{Lemma}[section]
\newcommand{\be}{\begin{equation}}
\newcommand{\ee}{\end{equation}}
\newcommand{\ba}{\begin{eqnarray}}
\newcommand{\ea}{\end{eqnarray}}
\newcommand{\spz}{\hspace{0.5cm}}
\newcommand{\virg}{\spz,\spz}
\newcommand{\pvirg}{\spz;\spz}

\def\d{\delta}
\newcommand{\D}{\Delta}
\newcommand{\T}{\Theta}
\def\t{\theta}

\newcommand{\la}{\lambda}
\newcommand{\p}{\partial}
\newcommand{\dx}{\partial_x}
\newcommand{\dt}{\partial_t}
\newcommand{\dla}{\partial_{\lambda}}

\newcommand{\tp}{\otimes}
\newcommand{\ptp}{\stackrel{\otimes}{,}}
\newcommand{\nn}{\nonumber}
\newcommand{\lt}{\left(}
\newcommand{\rt}{\right)}

\newcommand{\buno}{\mbox{\bf 1}}

\newcommand{\cC}{{\cal C}}

\newcommand{\cG}{{\cal G}}

\newcommand{\cL}{{\cal L}}

\newcommand{\cN}{{\cal N}}

\newcommand{\cP}{{\cal P}}

\newcommand{\cR}{{\cal R}}

\newcommand{\cV}{{\cal V}}
\newcommand{\cW}{{\cal W}}

\newcommand{\NP}[1]{Nucl.\ Phys.\ {\bf #1}}
\newcommand{\PL}[1]{Phys.\ Lett.\ {\bf #1}}

\newcommand{\CMP}[1]{Comm.\ Math.\ Phys.\ {\bf #1}}
\newcommand{\CPAM}[1]{Comm.\ Pure\ Appl.\ Math.\ {\bf #1}}

\newcommand{\PRL}[1]{Phys.\ Rev.\ Lett.\ {\bf #1}}
\newcommand{\MPL}[1]{Mod.\ Phys.\ Lett.\ {\bf #1}}

\newcommand{\LMP}[1]{Lett.\ Math.\ Phys.\ {\bf #1}}

\begin{document}
\sloppy
\renewcommand{\thefootnote}{\fnsymbol{footnote}}

\newpage
\setcounter{page}{1}

\vspace{0.7cm}
\begin{flushright}
DTP/99/65\\ EP/99/156\\
December 1999
\end{flushright}
\vspace*{1cm}
\begin{center}
{\bf  Hidden local, quasi-local and non-local Symmetries in Integrable Systems}\\
\vspace{1.8cm}
{\large D.\ Fioravanti $^a$\footnote{E-mail:
dfiora@sissa.it} and M.\ Stanishkov $^b$\footnote{E-mail:
stanishkov@bo.infn.it ; on leave of absence from I.N.R.N.E. 
Sofia, Bulgaria}}\\  \vspace{.5cm} $^a${\em I.N.F.N. Sez. di Trieste and
S.I.S.S.A.-I.S.A.S.\\      Via Beirut 2-4 34013 Trieste, Italy} \\ $^b${\em
Dipartimento di Fisica, Universita' di Bologna\\  Via Irnerio 46, 40126
Bologna, Italy} \\  \end{center}
\vspace{1cm}

\renewcommand{\thefootnote}{\arabic{footnote}}
\setcounter{footnote}{0}

\begin{abstract}
{\noindent The knowledge} of {\it non usual} and sometimes {\it hidden}
symmetries of (classical) integrable systems provides a very powerful
setting-out of solutions of these models. Primarily, the understanding and
possibly the quantisation of intriguing symmetries could give rise to deeper
insight into the nature of field spectrum and correlation functions in quantum
integrable models. With this perspective in mind we will propose a general
framework for discovery and investigation of local, quasi-local and non-local
symmetries in classical integrable systems. We will pay particular attention
to the structure of symmetry algebra and to the r\^ole of conserved
quantities. We will also stress a nice unifying point of view about KdV
hierarchies and Toda field theories with the result of obtaining a Virasoro
algebra as exact symmetry of Sine-Gordon Model.

\end{abstract}
\newpage

\resection{Introduction.}

What is usually defined as a 1+1 dimensional Integrable System is a classical
or quantum field theory with the property to have an infinite number of {\it
local integrals of motion in involution} (LIMI), among which the hamiltonian
(energy) operator . This kind of symmetry does not allow the determination of
the most intriguing and interesting features of a system because of its
abelian character. Instead, the presence of an infinite dimensional
non-abelian algebra could {\it complete} the abelian algebra giving rise to
the possibility of building its representations, {\it i.e.} the spectrum (of
the energy) and the spectrum of fields. We may name this non-commuting algebra
a spectrum generating algebra. In different models and in a mysterious way the
presence of this spectrum generating symmetry is very often connected to the
abelian one.   This is the case of the {\it simplest} integrable quantum
theories -- the two Dimensional Conformal Field Theories ({\bf 2D-CFT's}) --
their common crucial property being covariance under the infinite dimensional
Virasoro symmetry, a true spectrum generating symmetry. Indeed, the Verma
modules (highest weight representations) of this algebra classify all the
local  fields in 2D-CFT's and turn out to be reducible because of the
occurrence of  vectors of null hermitian product with all other vectors, the
so called  {\it null-vectors}. The factorization by the modules generated over
the  null-vectors leads to a number of very interesting algebraic-geometrical 
properties such as fusion algebras, differential equations for  correlation
functions, etc. .

Unfortunately this beautiful picture collapses when one pushes the system away
from criticality by perturbing the original CFT with some relevant local field
 $\Phi$:
\be
S=S_{CFT}+\mu\int d^2z\, \Phi(z,\bar{z}),
\label{pert}
\ee
and from the infinite dimensional Virasoro symmetry only the Poincar\'e subalgebra 
survives the perturbation. Consequently, one of the most important open
problems in two Dimensional Quantum Field Theories ({\bf 2D-IQFT's}) is
the  construction of the spectrum of the local fileds and consequently the 
{\it computation} of their correlation functions. Actually, the CFT possesses
a bigger ${\cW}$-like  symmetry and in particular it is invariant under an
infinite dimensional  abelian subalgebra of the latter \cite{SY}. With
suitable deformations, this  abelian subalgebra survives the perturbation
(\ref{pert}), resulting in the  so-called  LIMI. As said before, this symmetry
does not carry sufficient information, and in particular one cannot build the
spectrum of an integrable theory of type (\ref{pert}) by means of LIMI alone.

In the literature, there are several attempts to find spectrum generating 
algebras at least at conformal point. For instance in \cite{I}, 
some progress was made in arranging the spectrum generated through the action
of objects called {\it spinons} in representations of deformed algebras. The
kind of structure built by spinons eliminates automatically all null vectors
and consequently is unclear how to get (equations for) correlation functions.
Besides, the extension of this method in the scaling limit out of criticality
is up to now unknown. More recenlty, it has been conjectured in \cite{BBS}
that one could add to LIMI $I_{2m+1}$ additional non-commuting {\it charges}
$J_{2m}$ in such a way that the  resulting {\it algebra} (actually it is not
clear from \cite{BBS} if these objects close an algebra) would be sufficient
to create all the states of a particular class of perturbed theory (Restricted 
Sine-Gordon Theory) of type (\ref{pert}). Therein it
was also discovered that a sort of  null-vector condition appears in the above
procedure leading to certain equations for the form factors. However, what
remains unclear in \cite{BBS} are the field theory expressions of null
vectors, the general procedure for finding them and above all the symmetry
structures lying behind their arising. Besides, heavy use is
made of the very specific form of the form factors of the RSG model, and it is
 not clear how to extend this procedure to other integrable theories of 
the form (\ref{pert}). What is promising in this work is the link the authors
created between (quantum) Form Factors and classical solutions of the equation
of motion.

With this connection in mind, we present here a general framework to investigate 
symmetries and related charges in 2D integrable classical field theories. The
plan of the paper is as follows.

In Section 2 we present the central idea, {\it i.e.} basing the whole construction 
on a generalization of the so-called Dressing Symmetry Transformations
\cite{Sem} connecting these to the usual way of finding integrable systems
\cite{DS}. In fact, our basic objects  will be the {\it transfer matrix}
$T(x;\la)$, which generates the dressing,  and the {\it resolvent} $Z(x;\la)$,
the dressed generator of the underlying  symmetry. Although it is clear from
the construction that our method is applicable to any generalized  KdV
hierarchy \cite{DS}, we will be concerned with the semiclassical limit
\cite{GKM,BLZ,FRS} of minimal CFT's \cite{BPZ}, namely  the $A_1^{(1)}$-KdV
and the $A_2^{(2)}$-KdV systems \cite{DS}. Besides, we will show how to obtain
in a geometrical way from these conformal hierarchies the non conformal
Sine-Gordon Model (SGM), {\it i.e.} the semiclassical limit of Perturbed
Conformal Field Theories (PCFT's) (\ref{pert}). This kind of geometrical point
of view on the SGM (Toda field theories, in general), is very useful in
derivation of new symmetries and more general Toda field theories are linked
to generalized KdV equations.  

In Section 3 we will present an alternative approach to the description of 
the spectrum of the local fields in the classical limit of the 2D Integrable
Field Theories. We will build a systematic and geometric method for deriving
constraints or classical null vectors without the use of Virasoro algebra.

In Section 4 we will propose a further generalization of the aformentioned
dressing transformations. The central idea  is that we may dress not only the
generators of the underlying Kac-Moody algebra but also differential operators
in the spectral parameter,  $\lambda^m\dla^n$, forming a $w_\infty$ algebra.
The corresponding vector fields close a $w_\infty$ algebra as well with a
Virasoro subalgebra (realized for $n=1$) made up of quasi-local
and non-local transformations. The regular non-local ones are
expressed in terms of vertex operators and complete the quasi-local asymptotic
ones to the full Virasoro algebra. All these vector fields do not commute with
the KdV hierarchy flows, but have a sort of spectrum generating action on them.
Besides, it is very intriguing that only the positive index ones are exact
symmetries of the SGM (apparently the negative index ones do not matter
particularly in this theory). The apparition of a Virasoro symmetry, with its
rich and well known structure, is particularly useful and interesting.

In Section 5 we will deal with the  $A_2^{(2)}$-KdV showing, as an example of
generalization, the  building of the spectrum of local fields through the same
geometrical lines as before.  

In Section 6 we will summarize our results giving some hint on
the meaning and quantisation of these symmetries in critical and off--critical
theories.

\resection{The usual $A_1^{(1)}$--mKdV.}

\subsection{Introductory remarks on integrability.}
As already observed in  \cite{GKM}, the classical limit ($c\to
-\infty$) of CFT's is described by the second Hamiltonian structure of the
(usual) KdV which is built through the centerless Kac-Moody algebra $A_1^{(1)}$
in  the Drinfeld-Sokolov scheme \cite{DS}. In the classification  \cite{DS} a
generalized  modified-KdV (mKdV) hierarchy
is attached to each affine Kac-Moody loop algebra $\cG$. Various Miura
transformations \cite{Miu} relate it to the generalized KdV hierarchies, each one
classified by the choice of a node $c_m$ of the Dynkin diagram
of $\cG$. However, nodes symmetrical under automorphisms of
the Dynkin diagram lead to the same hierarchy. The
classical Poisson structure of this hierarchy is a classical
$w(\tilde{\cG})$-algebra, where $\tilde{\cG}$ is the finite dimensional Lie algebra
obtained by deleting the $c_m$ node. In the simplest case the usual
mKdV equation 
\be
\dt v=-\frac{3}{2}v^2v'-\frac{1}{4}v'''
\label{mkdv}
\ee
describes the temporal flow for the spatial derivative
\be
v=-\phi'
\ee
of a  Darboux field $\phi$ defined on a spatial interval $x\in [0,L]$. 
Assuming quasi-periodic boundary conditions on $\phi(x)$, it verifies by
definition the Poisson bracket 
\be
\{ \phi(x),\phi(y) \}={1\over 2}s(x-y) ,
\label{phi}
\ee
with the quasi-periodic extension of the sign-function $s(x)$ defined  as 
\be
s(x)=2n+1, \spz nL<x<(n+1)L .
\ee
As all the generalized mKdV, the simplest one (\ref{mkdv}) can be re-written 
as a null curvature condition \cite{Lax}
\be 
[ \dt - A_t , \dx - A_x ] = 0 
\ee 
for connections
belonging to the $A_1^{(1)}$ loop algebra 
\ba
A_x &=& - v h + (e_0 + e_1),  \nonumber \\
A_t &=& \la^2(e_0 + e_1 - vh) -
\frac{1}{2}[(v^2-v')e_0 + (v^2+v')e_1] - \frac{1}{2}(\frac{v''}{2}-v^3)h
\label{lax}
\ea
where the generators $e_0,e_1,h$  are chosen in the canonical gradation 
of the $A_1^{(1)}$ loop algebra
\be
e_0=\la E \virg e_1=\la F \virg h=H,
\ee
with $E,F,H$ generators of $A_1$ Lie algebra:
\be
[H,E]=2E \virg [H,F]=-2F \virg [E,F]=H.
\label{sl2}
\ee
For reasons of simplicity we will deal with the fundamental representation
\be
e_0=\left(\begin{array}{cc} 0 & \la \\
                                  0 & 0 \end{array}\right)\virg
e_1=\left(\begin{array}{cc} 0 & 0 \\
                                  \la & 0 \end{array}\right)\virg
h  =\left(\begin{array}{cc} 1 & 0 \\
                               0 & -1 \end{array}\right) .
\label{gen}
\ee
The KdV variable  $u(x,t)$ is related to the mKdV variable
$\phi(x)$  by the Miura transformation \cite{Miu} 
\be
u(x)=-\phi'(x)^2 -\phi''(x),
\label{miu}
\ee
which is the classical
counterpart of the quantum Feigin-Fuchs transformation \cite{FF}.
A remarkable geometrical interest is obviously attached to 
the transfer matrix which performs the parallel transport 
along the $x$-axis, {\it i.e.} the solution of the boundary value problem
\ba
\dx M(x;\la) &=& A_x(x;\la)M(x;\la) \nonumber \\
M(0;\la) &=& \buno . 
\label{M}
\ea
The formal solution of the previous equation 
\be
M(x,\la) ={\cP}e^{\int_0^xA_x(y,\la)dy}
\label{forsol}
\ee
can be expressed  by means of
a series of non-negative powers of $\la$ with an infinite convergence 
radius and non-local coefficients. The main property of the solution (\ref{forsol}) 
is that it allows one to calculate the equal time Poisson brackets between
the entries of the monodromy matrix  
\be
M(\la)=M(L;\la)={\cP}e^{\int_0^LA_x(y,\la)dy},
\label{mon}
\ee
provided those among the entries of the connection $A_x$
are known \cite{Fa84}. The result of this calculation is that the Poisson brackets 
of the entries of the monodromy matrix are fixed by the so called
classical $r$-matrix  \cite{Fa84}  
\be
\{ M(\lambda) \ptp M(\mu) \}=[r(\lambda\mu^{-1}) , M(\lambda) \tp M(\mu) ].
\label{PB's}
\ee
In our particular case the $r$-matrix is the trigonometric one (calculated, possibly, 
in the fundamental representation of $sl(2)$):
\be
r(\la)=\frac{\la+\la^{-1}}{\la-\la^{-1}}\frac{H\tp H}{2}+\frac{2}{ \la-\la^{-1}}
(E\tp F +F\tp E) .
\label{r}
\ee
By carrying through the trace on both members of the
Poisson brackets (\ref{PB's}) we are allowed to conclude immediately
that  
\be
\tau(\la)=trM(\la)
\label{tau}
\ee
Poisson-commute with itself for different values of the spectral parameter
\be
\{\tau(\la),\tau(\mu)\}=0.
\ee
In other words, $\tau(\la)$ is the generating 
function of the conserved charges in involution, 
{\it i.e.} it guarantees the integrability 
of the model {\it \`a la} Liouville. Now, it is possible to expand  the generating 
function $\tau(\la)$ in two different independent ways 
in order to obtain two different sets of conserved charges in involution. One is the regular
expansion in non-negative powers of $\la$, 
the other is the asymptotic expansion in negative powers. In the first case the coefficients 
in the Taylor series are non local charges of the theory, instead in the
second one the coefficients in the asymptotic series are the LIMI.
Likewise the transfer matrix $T(x,\la)$ can be expanded in the two  ways just
mentioned giving rise to different algebraic  and geometric structures, as we
will see in the following. The regular expansion  is typically employed  in
the derivation of Poisson-Lie  structures for Dressing Symmetries
\cite{Sem}.  Instead, the second type of expansion plays a  crucial r\^ole 
in obtaining the flows of the integrable hierarchy and the local integrals of motion
that generate symplectically these flows \cite{Fa84,DS}. 
In this section we will see how the aforementioned
approaches are actually reconducible to a single geometrical procedure which, moreover, 
produces two different kinds of symmetries.

\subsection{Regular expansion of the transfer matrix and the Dressing Transformations.}

The coefficients of the regular expansion of $\tau$ (\ref{tau}) are non-local integrals 
of motion in involution (NLIMI). However, these latter may be included in a larger
non-abelian algebra of  conserved charges, {\it i.e.} commuting with local
hamiltonians of the mKdV (\ref{mkdv}), but not all between themselves.
Actually to  get those in a suitable form, we use a slightly different
procedure than the usual one \cite{Sem}, considering a solution of the
associated linear problem (\ref{M}) with a different  initial condition.
Explicitly, we select the following  solution of the first equation in
(\ref{forsol}) which contains {\it the fundamental primary field} $e^{\phi}$
\cite{MS}: 
\be 
T_{reg}(x;\la)=e^{H\phi(x)}{\cal P} \exp\lt\la \int_0^xdy
(e^{-2\phi(y)} E+ e^{2\phi(y)} F ) \rt \label{regsol} \ee or, equivalently,
defining $K(x)= e^{-2\phi(x)}E+e^{2\phi(x)}F$, 
\be
T_{reg}(x;\la)=e^{H\phi(x)} \sum_{k=0}^{\infty}
\la^k \int_{x\geq x_1 \geq x_2 \geq ... \geq
x_k\geq 0}K(x_1)K(x_2)...K(x_k) dx_1 dx_2 ... dx_k .
\label{regexp}
\ee
Now we apply the usual dressing techniques \cite{Sem} using the previous expression 
$T_{reg}(x;\la)$ and stressing the point of contact with the derivation of an integrable
hierarchy given in \cite{DS}. If we define a {\it resolvent} $Z^X(x,\la)$ for
the Lax operator ${\cL}=\dx - A_x$   (\ref{lax}) as a solution of the equation
\be 
[{\cL},Z^X(x;\la)]=0  , \label{resdef}
\ee
it turns out that we may build several solutions by mean of a dressing reformulation of the first 
equation of (\ref{M}) 
\be
T\dx T^{-1}=\dx - A_x.
\ee
In the specific case of the regular expansion (\ref{regexp}), if $X=H,E,F$ is one of the generators 
(\ref{sl2}) we get a regular {\it resolvent} by dressing
\be 
Z^X(x,\la)=(T_{reg}XT_{reg}^{-1})(x,\la)=\sum_{k=0}^\infty \la^k Z^X_k  . 
\label{dres}
\ee

The definition (\ref{resdef}) of the resolvent is the key property for the construction of a 
symmetry algebra, since, once the gauge connection
\be
\Theta^X_n(x;\la)=(\la^{-n}Z^X(x;\la))_-=\sum_{k=0}^{n-1} \la^{k-n} Z^X_k  
\label{theta}
\ee 
is constructed, the commutator $[{\cL},\Theta^X_n(x;\la)]$ is in $\la$ of the
same degree of $[{\cL}, (\la^{-n}Z^X(x;\la))_+]$ and hence of degree zero.
Therefore, to get a self-consistent gauge transformation   
\be 
\Delta^X_n{\cL}
= [\Theta^X_n(x;\la),{\cL}] , \label{dresvec}
\ee
we have to require only that the r.h.s. in (\ref{dresvec}) is
proportional to $H$. This depends, for X fixed, on whether $n$ is even or
odd. Indeed, a recursive relation between the
terms $Z^X_n$ in (\ref{dres}) follows straightforward from the definition (\ref{resdef})
\ba
\dx Z^X_0 &=& \phi' [H,Z^X_0] \nonumber \\
\dx Z^X_n &=& \phi' [H,Z^X_n] + [E+F,Z^X_{n-1}] .
\label{recrel}
\ea
and allows us to find the modes $Z^X_n$ once the different {\it initial conditions} are established 
by inserting the first term of the expansion (\ref{regexp}) into (\ref{dres}) 
\be 
Z^H_0 = H     \virg       Z^E_0 = e^{2\phi}E     \virg    Z^F_0 = e^{-2\phi}F.
\label{inicon}
\ee
The previous two relations yield the various terms of the expansion of the resolvent in the form
\ba
Z^H_{2m}(x)=a^H_{2m}(x)H  \virg   Z^H_{2m+1}(x)=b^H_{2m+1}(x)E+c^H_{2m+1}(x)F \nonumber\\
Z^E_{2r}(x)=b^E_{2r}(x)E+c^E_{2r}(x)F  \virg  Z^E_{2r+1}(x)=a^E_{2r+1}(x)H  \nonumber \\
Z^F_{2p}(x)=b^F_{2p}(x)E+c^F_{2p}(x)F \virg  Z^F_{2p+1}(x)=a^F_{2p+1}(x)H 
\label{Z's}
\ea
where $a^X_n$,$b^X_n$,$c^X_n$ are non-local integral expressions, containing exponentials of the 
field $\phi$. In addition, the variation (\ref{dresvec}) may be explicitly
calculated as  
\be
\Delta^X_n A_x = [Z^X_{n-1},E+F] 
\label{expvar}
\ee
and hence it is clear that $Z^X_{n-1}$ cannot
contain any term proportional to $H$.   The conclusions about the parity of $n$ of (\ref{expvar}) 
are that
\begin{itemize}
\item in the $Z^H$ case, $n$ in (\ref{dresvec}) must be even,
\item in the  $Z^E$ and  $Z^F$ case, $n$ must conversely be odd. 
\end{itemize}

From (\ref{recrel}) it is possible to work out simple recursion relations for the coefficients 
$a^X_n$,$b^X_n$,$c^X_n$ in (\ref{Z's})
\ba
a'^X_n=c^X_{n-1}-b^X_{n-1}, \nonumber\\ 
b'^X_{n+1}-2\phi'b^X_{n+1}+2a^X_n=0, \nonumber\\
c'^X_{n+1}+2\phi'c^X_{n+1}-2a^X_n=0
\label{abcdres} 
\ea
where $n$ is even for $X=H$ and odd for $X=E,F$. In fact, another way
to obtain  this expansion could be to substitute directly the regular
expansion (\ref{regexp}) in (\ref{dres}), but the recursive relations (\ref{abcdres}) (with the 
initial conditions (\ref{inicon})) will provide a contact with the use of
{\it the recursive operator} \cite{Cal} in the theory of integrable
hierarchies. Now,  the action (\ref{expvar}) of the symmetry generators on the
bosonic field is given in terms of $a^X_n(x)$ by making use of (\ref{abcdres})
\be 
\D^X_n \phi'=-\dx a^X_n(x)  
\label{mkdvdres}
\ee
in which $n$ is even for $X=H$ and odd for $X=E,F$.
Instead, using the previous equations of motions and the relations (\ref{abcdres}) it is simple to 
show that the action on the {\it classical stress-energy tensor} $u(x)$
(\ref{miu}) is given in terms of $b^X_{n-1}(x)$ 
\be
\D^X_n u =-2\dx b^X_{n-1}(x) , 
\label{kdvdres}
\ee
in which $n$ is even for $X=H$ and odd for $X=E,F$. Now, we are interested in finding a recursive 
relation giving $a^X_n$ in terms of $a^X_{n+2}$ and vice versa. If we
indicate with $\dx^{-1}=\int^x_0dy$, the system of equations (\ref{abcdres})
yields easily 
\ba
a^X_n&=&\cR_a a^X_{n+2} \virg \cR_a=-v\dx^{-1}v\dx + {1\over 4}\dx^2,  \nn\\
a^X_{n+2}&=&\cR_a^{-1} a^X_n \virg \cR_a^{-1}=2(\dx^{-1}e^{-2\phi}\dx^{-1}e^{2\phi}+\dx^{-1}
e^{2\phi}\dx^{-1}e^{-2\phi}) ,
\label{dreca}
\ea  
in which $n$ is even for $X=H$ and odd for $X=E,F$.
Similarly, for $b^X_{n-1}$ and $b^X_{n+1}$
\ba
b^X_{n-1}&=&\cR_b b^X_{n+1} \virg \cR_b ={1\over 2}(u+\dx^{-1}u\dx+{1\over 2}\dx^2) , \nn\\
b^X_{n+1}&=&\cR_b^{-1}b^X_{n-1} \virg \cR_b^{-1} =2\dx^{-1}-{1\over 2}\cR_a^{-1}\dx  , 
\label{drecb}  
\ea
in which $n$ is even for $X=H$ and odd for $X=E,F$.
The linear differential operators $\cR_a$, $\cR_b$ are called recursive operators  \cite{Cal} and 
they generate the integrable flows of an hierarchy (next Section). We have
proven here that the proper dressing transformation
(\ref{mkdvdres}),(\ref{kdvdres}) can be thought of as generated by the
inverse power of the recursive operators, {\it i.e.}, in a compact notation, 
\be  
\D^X_n \phi' =-\dx\cR_a^{-{n-\nu}\over 2}a^X_{\nu} \virg \D^X_n u
=-2\dx\cR_b^{-{n-1-\nu'}\over 2}b^X_{\nu'}  \ee where $\nu(H)=0$,
$\nu(E)=\nu(F)=1$ and $\nu'(H)=1$, $\nu'(E)=\nu'(F)=0$.

Now we develop a general scheme to find the algebra of the infinitesimal dressing transformations 
(\ref{dresvec}) and we will use the same procedure to find the commutation
relations for the whole symmetry algebra we will discuss in the next sections.
The procedure is based on three steps.

\begin{lemma}
The equations of motion of the resolvents (\ref{dres}) under the flows (\ref{dresvec}) 
have the form :
\be
\Delta^X_n Z^Y=  [\Theta^X_n,Z^Y] -\la^{-n}Z^{[X,Y]}. 
\ee
\label{dreslem1}
\end{lemma}

   {\it Proof.} As first step we prove that 
\be
\tilde{Z}= \Delta^X_n Z^Y - [\Theta^X_n,Z^Y]
\label{zt}
\ee
is a resolvent (of ${\cL}$):
\ba
0&=&\Delta^X_n[{\cL},Z^Y]=[[\Theta^X_n,{\cL}],Z^Y]+[{\cL},\Delta^X_nZ^Y]= -[[Z^Y,\Theta^X_n],{\cL}]-
[[{\cL},Z^Y], \Theta^X_n]+ \nonumber  \\
&+&[{\cL},\Delta^X_nZ^Y]= [{\cL},\Delta^X_n Z^Y - [\Theta^X_n,Z^Y]].
\ea 
From definition (\ref{zt}) it has the form  
\be
\tilde{Z}=\sum_{l=0}^\infty \la^l(\d^X_nZ^Y_l-\sum_{m=0}^n[Z_m^X,Z_{l+n-m}^X])- 
\sum_{l=-n}^{-1} \la^{-l}Z^{[X,Y]}_{l+n} .
\ee
Now, we must distinguish two cases. In the first case $X\neq Y$ and the
first negative powers are 
\be
\tilde{Z}= -\la^{-r}Z^{[X,Y]}_0 - \dots .
\ee
But, given the first term of a resolvent, it is completely determined by the recursive 
relations (\ref{recrel}). In the second case $X=Y$ and the last term in
the previous equation  vanishes. Therefore $\tilde{Z}$ is expressed by the
series \be
\tilde{Z}=\sum_{l=0}^\infty \la^l(\d^X_nZ^Y_l-\sum_{m=0}^n[Z_m^X,Z_{l+n-m}^X])
\ee
and the first non-zero term must be a linear combination of $Z_0^H$,$Z_0^E$,$Z_0^F$ (see the first 
of equations (\ref{recrel})). It is clear that this is not 
possible and consequently $\tilde{Z}=0$, {\it q.e.m.}.

\begin{lemma}
The equations of motion of the connections (\ref{theta}) under the flows (\ref{dresvec}) 
have the form :
\be
\D^X_n\T^Y_s- \D^Y_s\T^X_n=[\T^X_n,\T^Y_s]-\T^{[X,Y]}_{n+s} . 
\ee
\label{dreslem2}
\end{lemma}

   {\it Proof.} By using the definition of $\T_n$ (\ref{theta}) and the previous Lemma \ref{dreslem1}, we obtain
\ba
\D^X_n\T^Y_s&-&\D^Y_s\T^X_n = \nn\\
&=&(\la^{-s}[\T^X_n,Z^Y])_- -(\la^{-n-s}Z^{[X,Y]})_- -(\la^{-n}[\T^Y_s,Z^X])_- 
+(\la^{-n-s}Z^{[Y,X]})_-= \nn\\
&=& -([\la^{-s}Z^Y,(\la^{-n}Z^X)_-])_-
-([(\la^{-s}Z^Y)_-,\la^{-n}Z^X])_--2\T^{[X,Y]}_{n+s}= \nn\\ &=&
([\la^{-s}Z^Y,(\la^{-n}Z^X)_+
-\la^{-n}Z^X])_--([(\la^{-s}Z^Y)_-,\la^{-n}Z^X])_-  -2\T^{[X,Y]}_{n+s}= \nn\\
&=& ([(\la^{-s}Z^Y)_+
+(\la^{-s}Z^Y)_-,(\la^{-n}Z^X)_+])_--([(\la^{-s}Z^Y)-,\la^{-n}Z^X])_- 
-\T^{[X,Y]}_{n+s} \nn\\ &=&
([(\la^{-s}Z^Y)_-,(\la^{-n}Z^X)_+])_--([(\la^{-s}Z^Y)-,\la^{-n}Z^X])_-
-\T^{[X,Y]}_{n+s}, \ea from which the claim follows very simply, {\it q.e.m.}.

\begin{theorem}
The algebra of the vector fields (\ref{dresvec}) form a representation of (twisted) Borel 
subalgebra $A_1\tp {\bf C}$ (of the loop algebra $A_1^{(1)}$):
\be
[\Delta^X_n , \Delta^Y_s] = -\Delta^{[X,Y]}_{n+s} \pvirg X,Y = H,E,F  .
\label{dressalg}
\ee
\label{dresth}
\end{theorem}
    
      {\it Proof.} We have to evaluate the action  of the commutator in the l.h.s. on 
${\cL}$ by using the equation of motion of ${\cL}$ (\ref{dresvec}), the
previous Lemma (\ref{dreslem2}) and the Jacobi identity: 
\ba 
[\Delta^X_n ,\Delta^Y_s]{\cL}&=&\D^X_n[\T^Y_s,{\cL}]-\D^Y_s[\T^X_n,{\cL}]= \nn\\
&=&[\D^X_n\T^Y_s-
\D^Y_s\T^X_n,{\cL}]+[\T^Y_s,[\T^X_n,{\cL}]]-[\T^X_n,[\T^Y_s,{\cL}]]= \nn\\
&=&-[\T^{[X,Y]}_{n+s},{\cL}], \ea
which is exactly the claim, {\it q.e.m.}.

To get from the previous Theorem \ref{dresth} the usual form of the algebra, it is enough 
to undertake the replacement $\D\rightarrow -\D$ and untwist. This kind of
transformations are historically called dressing transformations \cite{Sem}.
In consideration of the  fact that all our symmetries will be obtained by
dressing, we will call them {\it proper} dressing transformations (or flows).
In the case of mKdV, these flows are non local except the first ones which
have the form of a Liouville model equation of motion: 
\be
\Delta^E_1\phi'(x)=e^{2\phi(x)} \virg \Delta^F_1 \phi'(x) =  - e^{-2\phi(x)}.
\label{fcr} 
\ee 
In particular, the Theorem \ref{dresth} means that these
infinitesimal variations (\ref{fcr}) generate by successive commutations all
the proper dressing flows. In addition, from them it is simple to get the
Sine-Gordon equation in light-cone coordinates $x_\pm$ for the boson 
\be
\phi\rightarrow {i\over 2}\phi , 
\ee 
if we define 
\be
x_-=x \virg {\p\over \p x_+}={1\over 2i}(\D^E_1 + \D^F_1).
\label{sg1}
\ee
Indeed, it comes from (\ref{fcr}) that
\be
\p_+\p_-\phi=\sin\phi
\label{sg2}
\ee
where it has been defined $\p_\pm={\p\over \p x_\pm}$.

The currents, originating from this symmetry algebra, 
can easily be found by applying the transformations (\ref{mkdvdres}) 
to both members of the continuity equation (\ref{mkdv}) 
\be
J^X_{t,n}=\dx a^X_n(x) \virg J^X_{x,n}=\D^X_n(-\frac{1}{2}v^3-\frac{1}{4}v'').
\label{curr}
\ee
To the $J^X_{t,n}$ correspond the non-local charges 
\be
Q^X_n=\int_0^L J^X_{t,n}=a^X_n(L),  
\label{char}
\ee
which are not necessarily conserved (depending on the boundary conditions), 
due to non-locality. 

It is possible to verify by explicit calculations or from the 
Poisson brackets (\ref{PB's}) that the charges themselves close a (twisted)
Borel subalgebra $A_1\tp {\bf C}$ (of the loop algebra $A_1^{(1)}$).

It is interesting to note that the action by which these
charges generate the transformations (\ref{mkdvdres}) is not always symplectic,
but only in the case of the variations $\Delta^E_1,\Delta^F_1$. For instance
the following Poisson brackets   
\ba
\Delta^E_1v=\{Q^E_1,v\} \virg  \Delta^F_1v=\{Q^F_1,v\}  \nonumber \\
\Delta^H_2v=\{Q^H_2,v\} + Q^E_1\{Q^F_1,v\} - Q^E_1\{Q^F_1,v\} 
\label{poilie}
\ea
denote how in the first case the action is symplectic, while in the second it
is of Poisson-Lie type \cite{Sem}.

As a matter of fact, we will compute in the next Section how the transformations 
(\ref{dresvec}) act on $\dt - A_t$ (and the other higher time Lax operators
of the hierarchy), finding that they do as a gauge transformations.

\subsection{The Integrable Hierarchy and the Asymptotic Dressing.}

It is however well-known \cite{Fa84,DS} that besides the regular expansion of 
the transfer matrix an asymptotic expansion exists for the latter . Since this will play an essential role in
our construction, we  shall review a few important points in the procedure to
obtain the asymptotic expansion.  The main idea is to apply a gauge
transformation $S(x)$ on the Lax operator ${\cL}$ in such a way that its new
connection $D(x;\la)$ will be diagonal : 
\be
(\dx - A_x(x))S(x)=S(x)(\dx+D(x)).
\label{diag}
\ee
Because of the previous equation $T(x;\la)$ takes the form 
\be 
T(x;\la)=KG(x;\la)e^{-\int_0^x dy D(y)}
\label{asyexp1}
\ee
where we put $S=KG$ with 
\be
K =\frac{\sqrt{2}}{2}\left(\begin{array}{cc}  1 & 1 \\
                                              1 & -1 \end{array}\right),
\ee
while $G$ verifies the following equation
\be
\dx G + \tilde{A}_x G=GD  \virg    \tilde{A_x}=K^{-1}A_xK .
\label{ricc}
\ee
It is clear now that the previous equation can be solved by finding the 
asymptotic expansion for $D(x;\la)$  
\be
D(x;\la)=\left(\begin{array}{cc}  d_+ & 0 \\
                                   0   & d_- \end{array}\right)= \sum_{i=-1}^{\infty}\la^{-i}
d_i(x)H^i ,
\label{asyexp2}
\ee
and expressing the asymptotic expansion of $G(x;\la)$ in terms of off-diagonal matrices
\be
G(x;\la)=
       \left(\begin{array}{cc}  1 & g_+ \\
                                g_- & -1  \end{array}\right) = H +
\sum_{j=1}^{\infty}\la^{-j}G_j(x) ,  
\label{asyexp2.1} 
\ee 
where the matrices $G_j(x)$ are off-diagonal with entries
\be 
(G_j(x))_{12}=g_j(x) \virg (G_j(x))_{21}=(-1)^{j+1}g_j(x). 
\ee
In addition, the off-diagonal part, $g_j(x)$, can be separated obeying an equation of 
Riccati type. The latter is solved by a recurrence formula for the $g_j(x)$ 
\be 
g_1=-{v\over 2} \virg g_{j+1}=\frac{1}{2}(g'_j + v\sum_{k=1}^{j-1}g_{i-j}g_j). 
\ee 
In addition, it is simple to see that the diagonal part $d_j(x)$, $j>0$, is related to $g_j(x)$ by    
\be
d_j=(-1)^{j+1} v g_j ,
\label{asyexp3}
\ee
and  is given by $d_{-1}=-1$, $d_0=0$. Note that the $d_{2n}(x)$ are
exactly the charge densities (of the mKdV equation) resulting from the
asymptotic expansion of 
\be
\tau(\la)=trM(\la), \spz M(\la)=T(L,\la)G^{-1}(0,\la)K^{-1},
\ee
if we impose quasi-periodic boundary conditions on $\phi$.

On the other hand, it is likewise known \cite{DS} that the construction of the mKdV flows goes
through the definition of the asymptotic expansion of a resolvent   
\be 
Z^H(x,\la)=\sum_{k=0}^\infty \la^{-k} Z^H_k, \spz Z_0^H=H .  
\ee
defined through the following property
\be
[{\cL},Z^H(x;\la)]=0.
\label{hasyres} 
\ee
The previous equation may be translated into a recursive system of
differential equations for the entries of $Z^H(x;\la)$ and the solution turns
out to have the form \be 
Z^H_{2k}(x)=b_{2k}(x)E+c_{2k}(x)F \virg
Z^H_{2k+1}(x)=a_{2k+1}(x)H, 
\label{solasy} 
\ee 
where 
\ba
a_{2k+1}=\phi'b_{2k}-\frac{1}{2}b'_{2k} \nonumber \\
a_{2k+1}=\phi'c_{2k}+\frac{1}{2}c'_{2k} \nonumber \\ 
a'_{2k-1} = c_{2k}-b_{2k}.  
\label{abcasy} 
\ea 
In a way similar to what has been done for proper dressing transformation, we
build through  $Z^H$ the hierarchy of commuting mKdV flows defining the gauge
connections  
\be
\theta^H_{2k+1}(x;\la)=(\la^{2k+1}Z^H(x;\la))_+=\sum_{j=0}^{2k+1}
\la^{2k+1-j} Z^H_j(x), \spz k\in {\Bbb N}
\label{htheta}   
\ee
and their induced transformation  
\be 
\delta^H_{2k+1}A_x =-[\theta^H_{2k+1}(x;\la),{\cL}].
\label{mkdvf} 
\ee 
The form of $Z^H_{2k+1}$ given by equation (\ref{solasy}) imposes the self-consistency 
requirement $[\theta^H_n(x;\la),{\cL}]\propto H$ satisfied only for odd
subscript $n=2k+1$.

The action (\ref{mkdvf}) of the mkdV-flows on the bosonic field can be re-cast in terms 
of $a_{2k+1}$ by using the recursive system (\ref{abcasy}) 
\be
\d_{2k+1} \phi'=\dx a_{2k+1}.  
\label{mkdvf'}
\ee
Instead, using the previous equations of motion and the relations (\ref{abcasy}) it is 
simple to show that the action on the {\it classical stress-energy tensor}
$u(x)$ (\ref{miu}) is given in terms of $b_{2k}(x)$ 
\be
\d_{2k+1} u =2\dx b_{2k+2}.
\label{kdvf}
\ee
These relations have a form similar to that of proper dressing symmetries and
consequently  we are again interested in separating the recursive system
(\ref{abcasy}) into one single recursion relation for $a_{2k+1}$. It is simple
to show that the desired equation involves exactly the same recursion
operator $\cR_a$ of equation (\ref{dreca}): 
\be 
a_{2k+1}(x)=\cR_a a_{2k-1}(x) \virg
\cR_a=-v\dx^{-1}v\dx + {1\over 4}\dx^2. \label{areca}
\ee
This equation determines uniquely $a_{2k+1}(x)$, once the initial value of  $a_1(x)$ has 
been given. For a similar reason we obtain a recursive differential equation
for $b_{2k+2}(x)$   
\be
b'_{2k+2}(x)={1\over 2}u'b_{2k}+ub'_{2k}+{1\over 4}b'''_{2k}.
\label{difb}
\ee
This equation determines uniquely $b_{2k+2}(x)$, once the initial value of $b_0$ has been 
given. Indeed it implies
\be
b_{2k+2}=\cR_b b_{2k}, 
\label{arecb}  
\ee
where $\cR_b$ is the same as in equation (\ref{drecb}). The arbitrariness in the initial 
condition for  $a_{2k+1}$ and $b_{2k}$ will be fixed in the following using
the geometrical interpretation of the resolvent  (equation (\ref{hasydre})).
The recursive  operators $\cR_a$, $\cR_b$ \cite{Cal} generate the integrable
flows of the hierarchy as implied by (\ref{mkdvf},\ref{mkdvf'}):  
\be 
\d_{2k+1}\phi' =\dx\cR_a^k\phi' \virg \d_{2k+1} u =2\dx\cR_b^k\cdot 1 . \ee

Now, like in the previous Section, it is interesting to interpret this solution $Z^H$ to 
equation (\ref{hasyres}) as generated by dressing through the asymptotic
expansion of $T(x,\la)$ (\ref{asyexp1}),(\ref{asyexp2}),(\ref{asyexp3})   \be
Z^H(x,\la)=(THT^{-1})(x,\la).  
\label{hasydre} 
\ee 
The previous similarity transformation fixes the initial conditions
\be
a_1(x)=\phi' \virg b_0=1
\ee
throughout which all the other $a_{2k+1}$ and $b_{2k}$ can be determined via (\ref{arecb}) and  
(\ref{areca}). As a consequence of this fact the $b_{2k}$ are the densities of
the LIMI. Indeed, the differential relation (\ref{difb}) (or equivalently
(\ref{arecb})) coincides (up to a normalization factor of $u$) with that
satisfied by the expansion modes of the diagonal of the resolvent of the
Sturm-Liuoville operator $\dx-u$ \cite{GD}. The initial condition $b_0=1$
makes the $b_{2k}$ proportional to the aforementioned modes \cite{GD}.

The observation (\ref{hasydre}) makes evident the same geometrical origin of integrable 
hierarchies and of their  proper dressing symmetries. In addition it will
allow us in the sequel to build a more general kind of symmetries and find out
their algebra.  Indeed, the first generalization of (\ref{hasydre}) consists
in the construction of the flows deriving from the resolvents \be
Z^E(x,\la)=(TET^{-1})(x,\la) \virg Z^F(x,\la)=(TFT^{-1})(x,\la). 
\label{efasydre} \ee
Unlike the previous case, these resolvents possess an expansion in all the
powers of $\la$ \be
Z^E(x,\la)=\sum_{i=-\infty}^{+\infty} \la^{-i} Z^E_i \virg Z^F(x,\la)=
\sum_{j=-\infty}^{+\infty} \la^{-j} Z^F_j .
\label{efasyexp}
\ee    
In terms of the data (\ref{asyexp2}),(\ref{asyexp2.1}) of the asymptotic transfer matrix, 
they take the form
\be
Z^E(x;\la) =\frac{1}{2(1+g_+g_-)}e^{-2I(x)}\left(
\begin{array}{cc}  g_-^2-1   & (g_-+1)^2 \\
                  -(g_--1)^2 & 1-g_-^2 \end{array}\right)
\label{easyres}
\ee
and
\be
Z^F(x;\la) =\frac{1}{2(1+g_+g_-)}e^{2I(x)}\left(
\begin{array}{cc}  g_+^2-1 & -(g_+-1)^2 \\
                  (g_++1)^2 & 1-g_+^2 \end{array}\right) ,
\label{fasyres}
\ee
after defining the function
\be
I(x)=-{1\over 2}\int_0^x(d_-(y)-d_+(y))dy=\sum_{k=-1}^{\infty}\la^{-2k-1}\int_0^x d_{2k+1}(y)dy,
\ee 
which generates the LIMI once calculated in $x=L$. Now, it is easy one to
convince himself that  the entries of the resolvents
(\ref{easyres}),(\ref{fasyres}) admit an expansion in {\it all} the (positive
and negative) powers of $\la$, {\it i.e.} that the modes $Z^E_i$ and $Z^F_j$
of the series (\ref{efasyexp}) are made up of linear combinations of {\it all}
the three Lie algebra generators $E$,$F$,$H$. This implies the impossibility
to satisfy the self-consistency condition $\d{\cL} \propto H$. Nevertheless ,
we can go over this difficulty by defining two other resolvents, combinations
of the previous ones 
\be 
Z^+(x,\la)=(T(E+F)T^{-1})(x,\la) \virg
Z^-(x,\la)=(T(E-F)T^{-1})(x,\la).   \label{+-asydre}  
\ee    
By using the expressions (\ref{easyres}),(\ref{fasyres}), we obtain the
following formul\ae \hspace{0.1cm} for the entries of $Z^\pm$ 
\ba 
(Z^+)_{11}
&=&\frac{1}{2(1+g_+g_-)} [(g_-^2-g_+^2-2)\cosh 2I +(g_+^2-g_-^2)\sinh 2I], \nn
\\  
(Z^+)_{12}&=&\frac{1}{2(1+g_+g_-)} [(g_-^2-g_+^2+2g_-+2g_+)\cosh 2I -\nn \\
&-&(g_-^2+g_+^2-2g_--2g_++2)\sinh 2I], \nn \\
(Z^+)_{21}&=&\frac{1}{2(1+g_+g_-)} [(g_+^2-g_-^2+2g_-+2g_+)\cosh 2I +\nn \\
&+&(g_-^2+g_+^2-2g_-+2g_++2)\sinh 2I], \nn 
\\ (Z^+)_{22}&=&-(Z^+)_{11} ; 
\label{+asyres} 
\ea 
and 
\ba 
(Z^-)_{11} &=&\frac{1}{2(1+g_+g_-)}
[(g_-^2-g_+^2)\cosh 2I -(g_+^2+g_-^2-2)\sinh 2I], \nn \\
(Z^-)_{12}&=&\frac{1}{2(1+g_+g_-)} [(g_-^2+g_+^2+2g_--2g_++2)\cosh 2I -\nn \\
&-&(g_+^2-g_-^2-2g_--2g_+)\sinh 2I], \nn \\ 
(Z^-)_{21}&=&\frac{1}{2(1+g_+g_-)}
[-(g_+^2+g_-^2+2g_+-2g_++2)\cosh 2I +\nn \\ &+&(g_-^2-g_+^2-2g_--2g_+)\sinh
2I], \nn \\ 
(Z^-)_{22}&=&-(Z^-)_{11} . 
\label{-asyres} 
\ea 
If we assume to denote
by $(e)$ and $(o)$ series with only even and odd powers of $\la$ respectively,
we have  
\ba 
g_+g_-=(e) &\virg& g_+^2+g_-^2=(e) \virg g_+^2-g_-^2=(o), \nn \\
g_++g_-=(o) &\virg& g_+-g_-=(e) . \ea     Consequently, the parity of the
entries of $Z^\pm$ is given by \be
Z^+ =\left(\begin{array}{cc}  (e) & (o) \\
                              (o) & (e) \end{array}\right) \virg
Z^- =\left(\begin{array}{cc}  (o) & (e) \\
                              (e) & (o) \end{array}\right),
\ee
or equivalently by
\ba
Z^+=\sum_{i=-\infty}^{+\infty}\la^{-2i-1}(b^+_{2i+1}E+c^+_{2i+1}F)+ 
\sum_{i=-\infty}^{+\infty} \la^{-2i}a^+_{2i}H \nn \\ 
Z^-=\sum_{j=-\infty}^{+\infty} \la^{-2j}(b^-_{2i}E+c^-_{2i}F)+
\sum_{j=-\infty}^{+\infty} \la^{-2j-1}a^-_{2i+1}H  . 
\ea     
It follows that self-consistency requirement may now be satisfied and we are
allowed to define two new series of dressing transformations through the
connections   
\ba
\theta^+_{2i}(x;\la)=(\la^{2i}Z^+(x;\la))_+=\sum_{l=-\infty}^{2i} \la^{2i-l}
Z^+_l(x), \spz i\in {\Bbb Z} \nn \\
\theta^-_{2j+1}(x;\la)=(\la^{2j+1}Z^-(x;\la))_+=\sum_{l=-\infty}^{2j+1}
\la^{2j+1-l} Z^-_l(x), \spz j\in {\Bbb Z} \label{+-theta} 
\ea
which are no more finite sums. Finally, the following gauge transformations of the Lax 
operator ${\cL}$
\be 
\delta^+_{2i}A_x =-[\theta^+_{2i}(x;\la),{\cL}] \virg \delta^-_{2j+1}A_x =
-[\theta^-_{2j+1}(x;\la),{\cL}]
\label{+-mkdvf} 
\ee
yield this compact form for the additional mKdV flows 
\be
\d_{2i} \phi'=\dx a^+_{2i} \virg \d_{2j+1} \phi'=\dx a^-_{2j+1}.  
\label{+-mkdvf'}
\ee
These flows are complicated series in $x$ with quasi-local coefficients, so that 
it would be very difficult to find their commutation relations by direct
computation.  

Therefore, to find the algebra of these additional dressing transformations (\ref{+-mkdvf}), 
we use the previous procedure based on three steps. 
\begin{lemma}
The equations of motion of the resolvents (\ref{+-asydre}),(\ref{hasydre}) under 
the flows (\ref{+-mkdvf}) have the form :
\be
\d^X_n Z^Y=  [\t^X_n(x;\la),Z^Y] -\la^{n}Z^{[X,Y]}, 
\ee
where now $X,Y=H,E+F,E-F$.
\label{asylem1}
\end{lemma}

   {\it Proof.} We omit the specific proof because it can be carried out along the 
lines of the analogous Lemma (\ref{dreslem1}).

\begin{lemma}
The equations of motion of the connections (\ref{htheta}),(\ref{+-theta}) under the 
flows (\ref{+-mkdvf}) have the form :
\be
\d^X_n\t_s- \d^Y_s\t_n=[\t^X_n,\t^Y_s]-\t^{[X,Y]}_{n+s} . 
\ee
\label{asylem2}
\end{lemma}

   {\it Proof.} The proof is analogous to that of Lemma (\ref{dreslem2}).

\begin{theorem}
The algebra of the asymptotic dressing vector fields (\ref{mkdvf}),(\ref{+-mkdvf}) is:
\ba
[\d^H_{2k+1},\d^+_{2i}] &=& -2\d^-_{2k+2i+1}, \spz k\in {\Bbb N},\spz i\in
{\Bbb Z},  \nn \\ \vspace{0.5cm}
[\d^H_{2k+1},\d^-_{2j+1}] &=& -2\d^+_{2k+2j+2}, \spz j\in {\Bbb Z}, \nn \\
\vspace{0.5cm}
[\d^H_{2k+1},\d^H_{2l+1}] &=& 0, \spz l\in {\Bbb N}, \nn \\
\vspace{0.5cm}
[\d^+_{2i},\d^-_{2j+1}]   &=&  2\d^H_{2i+2j+1}, 
\label{asyalg}
\ea
where in the last relation we have defined $\d^H_{2k+1}=0$ if $k<0$.
\label{asyth}
\end{theorem}
    
      {\it Proof.} As in the proof of Theorem (\ref{dresth}), the action on 
${\cL}$ of the commutators in the l.h.s. can be calculated by using the
equation of motion of ${\cL}$ (\ref{+-mkdvf}),(\ref{mkdvf}), the previous
Lemma (\ref{asylem2}) and the  Jacobi identity, {\it q.e.m.}.

The previous Theorem \ref{asyth} proves that the KdV flows form an hierarchy (
they commute with each other). Besides, they are local and the Lemma \ref{asylem2}
ensures that each Lax connection transforms in a gauge way under a generic flow. This is
why we may attach a { \it time} $t_k$ to each flow $\d^H_{2k+1}$ and think to
each flow as a true symmetry of all the others. Instead, the additional
asymptotic flows $\d^+_{2i}$ and $\d^-_{2j+1}$ do not commute with the
hierarchy flows, but close an algebra in which they are (in some sense) {\it
spectrum generating symmetries}.

It is easy to prove that the proper dressing transformation are true symmetries of the 
hierarchy as well.
\begin{lemma}
The transformation of the resolvent (\ref{hasydre}) under the regular flows 
(\ref{dresvec}) and the evolution with the times $t_k$ of the regular
resolvents (\ref{dres}) have the same form of the hierarchy flow of ${\cL}$:  
\be 
\d^X_n Z^H=  [\T^X_n,Z^H] \virg \d^H_{2k+1}Z^X=
\t^H_{2k+1},Z^X],  \ee
where for the regular resolvents we have $X=H,F,F$.
\label{regasylem1}
\end{lemma}
\begin{lemma}
The mKdV flows (\ref{mkdvf}) act as gauge transformations on the connections 
(\ref{theta}) of the proper dressing flows:
\be
\d^H_{2k+1}\T_n- \D^X_n\t_{2k+1}=[\t^H_{2k+1},\T^X_n] . 
\ee
\label{asyreglem2}
\end{lemma}
\begin{theorem}
The proper dressing vector fields (\ref{dresvec}) commute with the mKdV flows (\ref{mkdvf}):
\ba
[\d^H_{2k+1},\D^X_n] &=& 0 . 
\ea
\label{asyregth}
\end{theorem}   
In particular, the previous Theorem \ref{asyregth} implies that the light cone evolution 
$\p_+$ commutes with all the KdV flows, {\it i.e.} a different way to say
that the KdV hierarchy is a symmetry of the light-cone SG. In particular, the
symmetry generator $\delta^H_{2k+1}$ maps, at infinitesimal level, solution of SG into
solution.   

In consideration of the fact that these theories are classical limits of CFT's
and  PCFT's, let us concentrate our attention on the phase spaces of mKdV and
KdV systems, i.e. those objects which at the quantum level constitute the {\it
spectrum of fields}.

\resection{The spectrum of fields in the $A^{(1)}_1$ framework.}
In the mKdV theory a local field is a polynomial in $v=-\phi'$ and its derivatives and 
the space spanned by these polynomials is the space (Verma module) of the
descendant of the identity. Instead, the action (simple product in the
classical theory) of these polynomials on a primary field $e^{m\phi}$
generates the space (Verma module) of the descendants of this primary field.
In our approach to the spectrum of this classical limit of CFT's, we propose
here to treat the gauge fields, {\it i.e.} the entries  of $Z^H$ as {\it
fundamental} fields.     Let us start by considering the composite fields
$a_{2n+1}$, $b_{2n}$, and $c_{2n}$  of (\ref{solasy}). In this Section we will
suppress the index H. The differential equations (\ref{abcasy}) tell us
immediately that not all of them are independent, we may for example express
the $c_{2n}$ in terms of the {\it basic} fields $b_{2n}$ and $a_{2n+1}$. We
use now Lemma \ref{asylem1}   \be \delta_{2k+1}Z=[\theta_{2k+1},Z]
\label{moteqs} \ee which allows us to establish the action of each
$\delta_{2k+1}$ on these fields 
\ba
\delta_{2k+1}a_{2n+1}=\sum_{i=0}^{n}(a'_{2n+2k-2i+1}b_{2i}-a'_{2i-1}b_{2m+2k-2i+2}), \nonumber \\   
\delta_{2k+1}b_{2n}=2\sum_{i=0}^{n-1}(a_{2n+2k+1}b_{2n+2k-2i}-a_{2n+2k-2i+1}b_{2i})   
\label{deltaab}
\ea
Therefore, according to our conjecture the linear generators of the mKdV identity Verma module 
${\cV}^{mKdV}_{\buno}$ are made up of the repeated actions of the
$\delta_{2k+1}$ on the polynomials ${\cP}(b_2,b_4,\dots ,b_{2N},a_1,a_3,\dots
,a_{2P+1})\}$ in  the $b_{2n}$ and the $a_{2k+1}$   
\be
{\cV}^{mKdV}_{\buno}=\{linear\spz combinations\spz of \spz \delta_{2k_1+1}\delta_{2k_2+1}\dots 
\delta_{2k_M+1}{\cP}\} ,  
\label{veridm}
\ee
with a natural gradation provided by the subscripts. Actually, the Verma
module  ${\cV}^{mKdV}_{\buno}$ exhibits several null
vectors, i.e. polynomials in the $b_{2n}$ and the $a_{2k+1}$ 
which are zero. This is due to the very simple constraint on $Z^H$
\be
(Z^H)^2=\buno 
\label{con}
\ee
originating from the dressing relation with the transfer 
matrix $T$ (\ref{hasydre}). The constraints (\ref{con}) may be rewritten through the modes of 
$a_{2k+1}$ and $b_{2k}$    
\be
{\cC}_{2n}=\sum_{i=0}^nb_{2n-2i}(b_{2i}+a'_{2i-1})+\sum_{i=0}^{n-1}a_{2n-2i-1}a_{2i+1}=0,
\label{consab}
\ee
and produce null-vectors under the application of mKdV flows $\d_{2k+1}$. These latter generate 
linearly the graded vector space (Verma module) of all null vectors 
\be
{\cN}=\{linear \spz combinations\spz of\spz \delta_{2k_1+1}\delta_{2k_2+1}\delta_{2k_3+1} \dots 
\delta_{2k_Q+1}{\cC}_{2n}\}.
\label{null}
\ee
In conclusion our conjecture  is that the ({\it conformal}) family of the identity 
$[\buno]^{mKdV}$ of the mKdV hierarchy is obtained as a factor space:
\be 
[\buno]^{mKdV}={\cV}^{mKdV}_{\buno}/{\cN}.
\label{conidm}
\ee
On the other hand, in order to deduce the form of the Verma module
${\cV}^{KdV}_{\buno}$ of the identity for the KdV hierarchy we have to make three observations:
\begin{enumerate}
\item the recursive formula (\ref{arecb}) proves that $b_{2n}$ are polinomials of the KdV 
field $u(x)$ and its derivatives, whereas the $a_{2k+1}$ do not enjoy this
property;  \item the variation of $b_{2n}$ in (\ref{deltaab}) can be written
{\it accidentally} in
terms of the $b_{2k}$ alone, using the relationships (\ref{abcasy}) between 
$a_{2k+1}$ and $b_{2k}$
\be
\delta^H_{2k+1}b_{2n}=\sum_{i=0}^{n-1}(b'_{2n+2k-2i}b_{2i}-b'_{2i}b_{2m+2k-2i});
\label{deltab}
\ee
\item also the null vector space ${\cN}$ can be spanned by the $b_{2k}$
alone
\be
{\cC}_{2n}= b_{2n}+\sum_{i=1}^n[b_{2n-2i}b_{2i}-2b_2b_{2n-2i}b_{2i-2}-\frac{1}{2}
b_{2n-2i}b'_{2j-2}+\frac{1}{4}b'_{2n-2j}b'_{2j-2}] = 0 ,
\label{constb}
\ee 
using the relationships (\ref{abcasy}) between $a_{2k+1}$ and $b_{2k}$.
\end{enumerate}

Therefore, we conjecture that the Verma module  ${\cV}^{KdV}_{\buno}$ shall be linearly 
generated by elements given by repeated actions of the $\delta_{2k+1}$ on the
polynomials ${\cP}(b_2b_4\dots b_{2N})$ in the $b_{2n}$   
\be
{\cV}^{KdV}_{\buno}=\{linear \spz combinations\spz of\spz \delta_{2k_1+1}\delta_{2k_2+1} 
\dots \delta_{2k_M+1}{\cP}\} .
\label{veridk}
\ee
It turns out to be a sort of {\it reduction} of the 
Verma module ${\cV}^{mKdV}_{\buno}$ (\ref{veridm}) of the mKdV hierarchy. 
As for the mKdV case the ({\it conformal}) family of the identity 
$[\buno]^{KdV}$ of the KdV hierarchy is obtained as a factor space 
of ${\cV}^{KdV}_{\buno}$ over ${\cN}$ given by (\ref{null}) and 
(\ref{constb}): 
\be 
[\buno]^{KdV}={\cV}^{KdV}_{\buno}/ {\cN}.
\label{conidk}
\ee
Therefore we are led to the same scenario that arises
also in the classical limit of the construction \cite{BBS}. Nevertheless, in 
our approach the generation of null-vectors is automatic and geometrical (see equations  
(\ref{deltab}) and  (\ref{constb})). In addition , our approach is applicable to any other 
integrable system, based on a Lax pair formulation. We will illustrate this fact below
by using the example of the  $A_2^{(2)}$-mKdV system.
Other local fields of the mKdV system are the primary fields, i.e. 
the exponential $e^{m\phi}, m=0,1,2,3,\dots$ of the bosonic field. 
Indeed, for $m=0$ we obtain just the identity $\buno$, the 
{\it fundamental primary field} $e^{\phi}$ (m=1) appears in the 
regular expansion (\ref{regexp}) of the transfer matrix $T(x;\la)$ and the 
other primary fields $e^{m\phi}, m>1$ are the ingredients of the regular expansion 
of the power $T^m(x;\la)$. The previous construction of the identity operator 
family suggests the following form for the Verma module  ${\cV}^{mKdV}_m$ of 
the primary $e^{m\phi}, m=0,1,2,3,\dots$:  
\be
{\cV}^{mKdV}_m=\{linear \spz combinations\spz of\spz \delta_{2k_1+1}\delta_{2k_2+1}\dots 
\delta_{2k_M+1}[{\cP}e^{m\phi}]\} ,  
\label{vermmk}
\ee
where ${\cP}(b_2,b_4,\dots ,b_{2N},a_1,a_3,\dots ,a_{2P+1})$ are polinomials in the  $b_{2n}$ 
and the $a_{2k+1}$. As for the identity family, we have to subtract all the
null-vectors  (\ref{deltaab})  and  (\ref{constb}). Besides, in this case, we
have to take into account the null-vectors coming from the equations of motion
of the power  $T_{reg}^m(x;\la)$ of the regular  expansion
\be
\delta_{2k+1}T_{reg}^m=\sum_{j=1}^mT_{reg}^j\theta_{2k+1}T_{reg}^{m-j}.
\label{eqmT^m}
\ee
By successive applications of   (\ref{deltaab}), (\ref{constb}) and (\ref{eqmT^m}) 
we obtain the whole
null-vector set ${\cN}^{KdV}_{{\bf m}}$. In conclusion the spectrum is again the factor space 
\be
[{\bf m}]={\cV}^{mKdV}_m / {\cN}^{KdV}_{{\bf m}} .
\label{conmk}
\ee
Similarly, the construction of the identity operator 
family suggests the following form for the Verma module  ${\cV}^{KdV}_m$ of 
the primary $e^{m\phi}, m=0,1,2,3,\dots$:  
\be
{\cV}^{KdV}_m=\{linear \spz combinations\spz of\spz\delta_{2k_1+1}\delta_{2k_2+1}\dots 
\delta_{2k_M+1}[{\cP}(b_2,b_4,\dots ,b_{2N})e^{m\phi}]\} ,  
\label{vermkd}
\ee
with ${\cP}(b_2,b_4,\dots ,b_{2N})$ polinomials in $b_{2n}$. Again, by successive applications of  
(\ref{deltab}), (\ref{constb}) and (\ref{eqmT^m}) we obtain the whole
null-vector linear space ${\cN}^{KdV}_{{\bf m}}$. In conclusion, the spectrum
is again  a factor space  
\be
[{\bf m}]={\cV}^{KdV}_m / {\cN}^{KdV}_{{\bf m}} .
\label{conkd}
\ee
Of course, we have checked all our conjectures up to high gradation of the
null vectors.  Nevertheless, we did not manage to generate the spectrum of
fields only through the asymptotic symmetry of Theorem \ref{asyth}. For
dimensional arguments it is plausible to make the substitution 
\ba
&a_{2k+1}&\rightarrow \d^+_{2k+1}, \spz k\in {\Bbb Z} \nn \\
\vspace{0.5cm}
&b_{2k}&\rightarrow \d^-_{2k},
\ea
but now the null vector meaning and origin should be completely different.

\resection{A non local Virasoro symmetry by dressing.}

At this point we are in a position to construct in a natural way more general kinds of 
dressing-like symmetries. It is well known that the vector fields
$l_m=\lambda^{m+1}\dla$ on the circumpherence realize the centerless Virasoro
algebra  \be [l_m,l_n]=(m-n)l_{m+n}.
\ee 
A very natural dressing is represented by the resolvents
\ba
Z^V_{m}=T_{reg}l_{m}T^{-1}_{reg} , \spz m<0 \nn \\
Z^V_{m}=T_{asy}l_{m}T^{-1}_{asy} ,\spz   m\geq 0
\label{vr}
\ea 
where we have to use the different regular and asymptotic tranfer matrices,  $T_{reg}$ 
and $T_{asy}$. Of course, they satisfy the usual definition of  resolvent
\be
[{\cL},Z^V_{m}(x;\la)]=0
\label{vresdef}
\ee
and, as in the previous cases, they have two different kinds of expansions 
\ba
(Z^V_{-1})_{reg}=T_{reg}l_{-1}T^{-1}_{reg}=\sum_{n=0}^{\infty}\la^n Z^{reg}_{n+1} -\dla \nn \\
(Z^V_{-1})_{asy}=T_{asy}l_{-1}T^{-1}_{asy}=\sum_{n=0}^{\infty}\la^{-n} Z^{asy}_{n-1} -\dla
\label{vre}
\ea
and consquently the mode expansion of the more general Virasoro resolvent (\ref{vr}). In the 
same way, (\ref{vresdef}) authorizes us to define gauge connections 
\ba
\theta^V_m=(Z^V_m)_- =\sum_{n=0}^{-m-2}\la^{n+1+m}Z^{reg}_{n+1} -\la^{m+1}\dla, \spz m<0 \nn \\
\theta^V_{m}=(Z^V_{m})_+=\sum_{n=0}^{m+1}\la^{m+1-n}Z^{asy}_{n-1} -\la^{m+1}\dla,\spz m\geq 0
\label{vircon}
\ea
and the relative gauge transformations
\be
\delta_m^V A_x =-[\theta^V_{m}(x;\la),{\cL}] .
\label{vf}
\ee
Finally, we have to verify the consistency of this gauge transformation  requiring 
$\delta_m^V A_x=H \delta_m^V  \phi'$ for positive and negative $m$.
It is very easy to see that this requirement imposes $m$ to be even. Indeed, from 
(\ref{vresdef}) or (\ref{vr}) it is simple to derive the form of the generic
term of the expansions (\ref{vre}) 
\ba
Z^{reg}_{2n-1}=b^V_{2n-1}E+c^V_{2n-1}F &,&\spz  Z^{reg}_{2n}=a^V_{2n}H,\spz n>0 \nn \\
Z^{asy}_{2n-3}=\beta^V_{2n-3}E+\gamma^V_{2n-3}F &,&\spz  Z^{asy}_{2n-2}=\alpha^V_{2n-2}H,\spz n>0.
\ea
In addition, we can easily find recursive relations for the regular coefficients
\be
b^V_{2k+1}(x)=\cR^{-1}_b b^V_{2k-1}(x) \virg a^V_{2n+2}(x)=\cR^{-1}_a a^V_{2n}(x),
\ee
and for the asymptotic coefficients
\be
\beta^V_{2n-1}(x)=\cR_b\beta^V_{2n-3}(x) \virg \alpha^V_{2n}(x)=\cR_a
\alpha^V_{2n-2}(x), \label{avrr}
\ee
where the recursive operators $\cR_a$,$\cR_b$ are given in (\ref{dreca}),(\ref{drecb}) and 
(\ref{vr}) fixes the initial conditions:
\ba
&b^V_{-1}&=e^{2\phi}\dx^{-1}e^{-2\phi} \virg a_2=\int_0^x dx_2 \int_0^{x_2}dx_1
2\cosh[\phi(x_1)-\phi(x_2)], \nn \\
&\beta^V_{-1}&=x \virg \alpha^V_0=x\phi' . 
\ea
In conclusion the Virasoro mKdV flows are given by:
\ba
\d^V_{2m} \phi'=-\dx a^V_{-2m},\spz m<0 \nn \\ 
\d^V_{2m} \phi'=\dx \alpha^V_{2m},\spz m\geq 0,   
\label{vmkdvf'}
\ea
and explicit examples of the first flows are
\ba
\delta_{-2}^V \phi' &=&  e^{2\phi(x)} \int_0^x dy e^{-2\phi(y)} -  e^{-2\phi(x)} \int_0^x dy 
e^{2\phi(y)}=e^{2\phi(x)}B_1-e^{-2\phi(x)}C_1 \nonumber  \\ 
\delta_{-4}^V  \phi' &=&  e^{2\phi(x)}(3B_3(x)-A_2(x)B_1(x)) - 
e^{-2\phi(x)}(3C_3(x)-D_2(x)C_1(x)) \nonumber  \\
\delta_{-6}^V  \phi' &=& e^{2\phi(x)} (5B_5(x)-3A_4(x)B_1(x)+A_2(x)B_3(x)) \nonumber  \\
&-& e^{-2\phi(x)}(5C_5(x)-3D_4(x)C_1(x)+D_2(x)C_3(x)) \nn \\
\delta_{0}^V \phi' &=& \phi' + x\phi'' \nonumber \\
\delta_{2}^V  \phi' &=& 2xa_3'+6g_3-2g_1^3+2g'_1\int_0^x d_1,  
\label{frd's} 
\ea
where we have used a convenient notation for the entries of the regular expansion 
\be
T_{reg}(x,\la) =\left(\begin{array}{cc} A &  B \\
                                        C  &  D\end{array}\right) ,
\ee
with $A=e^{\phi}(1+\sum_1^{\infty}\lambda^{2n}A_{2n})$, $B=e^{\phi}\sum_0^{\infty}
\lambda^{2n+1}B_{2n+1}$, $C[\phi]=B[-\phi]$ and $D[\phi]=A[-\phi]$. We stress
that negative subscript variations have a form very similar to that of the
regular dressing flows ((\ref{dresvec}) with $X=H$) $\D_{2r}^H$.
Nevertheless, in spite of the commutativity $[\D_{2r}^H,\D_{2s}^H]=0$ we will
see that they obey instead Virasoro commutation relations. From the actions
(\ref{frd's}) the transformations of the classical {\it  primary fields} 
$e^{\phi}$ follow. For example: 
\ba 
\delta_{-2}^V e^{\phi} &=& (D_2-A_2)e^\phi
 \nonumber  \\ \delta_{-4}^V e^{\phi} &=& 
[(3D_4-C_3B_1)-(3A_4-B_3C_1)]e^\phi, \nn \\ \delta_{0}^V e^{\phi} &=&
(x\dx+\Delta)e^\phi  \nonumber  \\ \delta_{2}^V e^{\phi} &=&
(2xa_3+2g_2+2g_1\int_0^x d_1)e^\phi .   \ea It is understood of course that
these fields are primary with respect to the usual space-time Virasoro
symmetry. The actions of the variations (\ref{vmkdvf'}) on the generator of
this symmetry can be easily calculated as usual by means of Miura
transformation (\ref{miu}) and the recursive relations implied by
(\ref{vresdef})  
\ba 
\d^V_{2m} u=-2\dx b^V_{-2m-1}, \spz  m<0 \nn \\ 
\d^V_{2m} u=2\dx \beta^V_{2m+1}, \spz m\geq 0.    \label{vkdvf} \ea For example
\be
\delta^V_2 u= x(u'''-{3\over 2}uu')+u''-2u^2-{1\over 2}u'\int_0^xu . 
\ee

We now apply our usual procedure in three parts to find transformation equations and 
symmetry algebra in this case. We will omit the proofs in consideration of
the fact that they are very similar to the previous ones. \begin{lemma}
The equations of motion of the resolvents (\ref{vr}) under the flows (\ref{vf}) have the form :
\be
\delta_{2n}^VZ_{2m}^V= [\theta^V_{2n},Z^V_{2m}]-(2n-2m)Z^V_{2n+2m}, \spz m,n\in{\Bbb Z}. 
\ee
\label{virlem1}
\end{lemma}

\begin{lemma}
The transformations of the connections (\ref{vircon}) under the flows (\ref{vf}) have the form :
\be
\delta_{2n}^V\theta_{2m}^V-\delta_{2m}^V\theta_{2n}^V = [\theta^V_{2n},\theta^V_{2m}]-(2n-2m)
\theta^V_{2n+2m}, \spz m,n\in{\Bbb Z} . 
\ee
\label{virlem2}
\end{lemma}

\begin{theorem}
The algebra of the vector fields on ${\cL}$ (\ref{vf}) forms a representation of the centerless 
Virasoro algebra:
\be
[\delta^V_{2m},\delta^V_{2n}]=(2m-2n)\delta^V_{2m+2n}, \spz m,n\in{\Bbb Z},
\label{vir}
\ee
\label{virth}
after the redefinition $\d^V\rightarrow -\d^V$.
\end{theorem}

The Theorem \ref{virth} gives us a very non-trivial information because of
the different  character of the asymptotic and regular Virasoro vector fields.
Indeed, the asymptotic ones are quasi-local (they can be made local after
differentiating a certain number of times), the regular ones instead are
essentially non-local being expressed in terms of vertex operators. In
addition, it is easy to compute the most simple relations
$[\delta_0,\delta_{2n}]=-2n\delta_{2n}$, $n\in {\Bbb Z}$, which means that
$\delta_0$  counts the dimension or level. We want to stress once more that
this Virasoro symmetry is different from the space-time one and is essentially
non-local. The additional symmetries coming from the regular dressing are very
important for applications. They  complete the asymptotic ones forming an
entire Virasoro algebra and provide a possibility of a central extension in
the (generalized) KdV hierarchy, which is the classical limit of CFT's .
However, this central term may appear only in the algebra of the  hamiltonians
of the above transformations, as it is for the case of CFT' s as well.

With the aim of understanding the classical and quantum structure of integrable systems, 
we present here the complete algebra of symmetries. The Virasoro flows
commute neither with the mkdV hierarchy (\ref{mkdvf}) nor with the (proper)
regular dressing flows (\ref{dresvec}). In fact one can show, following the
lines of the three steps procedure, these statements. \begin{lemma} The
equations of motion of the resolvents (\ref{vr}) under the mKdV flows 
(\ref{mkdvf}) and of the mKdV resolvent (\ref{hasydre}) under the Virasoro
flows (\ref{vf}) have the same form of the variation of ${\cL}$: 
\ba
\delta^H_{2k+1}Z_{2m}^V&=& [\theta^H_{2k+1},Z^V_{2m}], \spz k\in{\Bbb N}, \spz
m\in{\Bbb Z} \nn \\  \delta_{2m}^VZ^H_{2k+1}&=& [\theta^V_{2m},Z^H_{2k+1}].
\ea \label{vmlem1} \end{lemma}

\begin{lemma}
The mixed transformations of the connections (\ref{vircon}) under the mKdV
flows (\ref{mkdvf}) are not of gauge type: 
\be
\delta^H_{2k+1}\theta_{2m}^V-\delta_{2m}^V\theta_{2k+1}=
[\theta^H_{2k+1},\theta^V_{2m}]-(2k+1)\theta^H_{2k+2m+1} .  \ee \label{vmlem2}
\end{lemma}

\begin{theorem}
The algebra of the hierarchy flows and of the Virasoro flows is not abelian:
\be
[\delta^H_{2k+1},\delta^V_{2m}]=(2k+1)\delta^H_{2m+2k+1},
\label{vm}
\ee
where we have put $\d^H_{2k+1}=0$ if $k<0$ in the r.h.s..
\label{vmth}
\end{theorem}

The content of the previous theorem is that Virasoro symmetry shifts along the
KdV hierarchy. 

Likewise, it may be proven that
\be
[\D^X_n,\delta^V_{2m}]= -n \D^X_{n-2m}.
\ee
after putting the proper dressing flows
(\ref{dresvec}) with negative $n-2m$: $\D^X_{n-2m}=0$  in the r.h.s.. As a very important
consequence of this fact, the light-cone Sine-Gordon flow $\p_+$
(\ref{sg1}),(\ref{sg2}) commutes  with all the positive Virasoro modes, {\it i.e.} we have 
obtained a half Virasoro algebra as exact symmetry of SGM. We note again that
this infinitesimal transformation is quasi-local in the boson $\phi$.

One remark is necessary at this stage. It is quite interesting to have a Virasoro algebra not 
commuting (spectrum generating) with the KdV flows, but one may transform the
Virasoro flows into true symmetries commuting with the mKdV hierarchy by
adding a term containing all the times $t_{2k+1}$  
\be
\delta^V_{2m} \rightarrow \delta^V_{2m} - \sum_{k=1}^{\infty}(2k+1)t_{2k+1}\delta^H_{2m+2k+1},
\ee
From the view point of CFT it is very difficult to give a physical meaning to these times, 
but from the restriction of the action of the positive part of the Virasoro
algebra on $u$ (\ref{vkdvf}) we can check that the previous formula yields the
half Virasoro algebra described in \cite{OS},\cite{GO} by using the
pseudodifferential operator method. Actually, it plays an important role in
the study of the matrix models where it leads to the so called Virasoro
constraints: $L_m\tau=0, m>0$. Here $\tau$ is the  $\tau$-function of the
hierarchy and is connected to the partition function of the matrix model.
Moreover, it seems that it should play an important role also in the context
of the Matrix String Theory \cite{YM}, which is now intensively studied. Note
also that these Virasoro constraints are the conditions for the highest
weight state and, because we also have $L_0\tau\propto \tau$ \cite{GO}, the
$\tau$-function is a primary state for the Virasoro algebra. But, we uncovered
the negative modes of the Virasoro algebra, which build the highest weight
representation over the $\tau$-function. We are analyzing this intruiging
scenario even in off-critical theories like Sine-Gordon \cite{35}.

Let us also note that actually the symmetry  of mKdV is much larger. Indeed, 
the differential operators $l_{2m,2n}=\lambda^{2m+1}\partial_\lambda^{2n+1}$
close a (twisted) $w_\infty$ which is isomorphic to its dressed version 
\be
\delta_{2m,2n} A_x =-[\theta_{2m,2n}(x;\la),{\cL}],
\label{wf}
\ee
where we have defined the connections and the resolvents
\ba
\theta_{2m,2n}=(Z^V_{2m,2n})_-,\spz Z_{2m,2n}=T_{reg}l_{2m,2n}T^{-1}_{reg},
\spz m<0 \nn \\ 
\theta_{2m,2n}=(Z^V_{2m,2n})_+,\spz
Z_{2m,2n}=T_{asy}l_{2m,2n}T^{-1}_{asy}, \spz  m\geq 0. 
\label{wcon}
\ea
In particular from the commutations of the {\it diagonal } differential
operators  $\l_{2n,2n}=\lambda^{2n+1}\partial_\lambda^{2n+1}$ 
\be
[\l_{2n,2n},\l_{2m,2m}]=0
\ee
we deduce the existence of a quasi-local hierarchy of the diagonal flows
\be
[\d_{2n,2n},\d_{2m,2m}]=0, \spz n,m>0.
\ee
As far as we know, this observation is new and we suggest that the diagonal flows are 
connected to the higher Calogero-Sutherland hamiltonian flows in their
collective field theory description. This could give a geometrical explicit
explanation of the misterious connection between Calogero-Sutherland systems
and KdV hierarchy \cite{AMM}.

\resection{Generalization: the $A_2^{(2)}$-KdV.}

Let us show that our approach is easily applicable to other integrable systems. Here we 
consider the case of the $A_2^{(2)}$-KdV equation. The reason is that it can be considered
as a different classical limit of the {\bf CFT's} \cite{FRS,PhD}. Consider the
matrix  representation of the $A_2^{(2)}$-KdV equation:
\be
\dt{\cL}=[{\cL}, A_t]
\label{mKdV2}
\ee
where
\be
{\cL}=\dx-A_x \virg A_x= \phi' h + (e_0 + e_1),
\label{a22lax} 
\ee
and 
\be
e_0= \left(\begin{array}{ccc} 0 & 0 & \la \\
                                  0 & 0 &  0  \\                                      
                                  0 & 0 &  0   \end{array}\right)\virg
e_1= \left(\begin{array}{ccc} 0 & 0 &  0 \\
                                  \la & 0 &  0  \\
                                  0 & \la & 0 \end{array}\right)\virg
h  = \left(\begin{array}{ccc} 1 & 0 & 0 \\
                               0 & 0 & 0 \\
                               0 & 0 & -1 \end{array}\right) 
\label{gen2}
\ee
are the generators of the Borel subalgebra of $A_2^{(2)}$ and $A_t$ is a 
certain connection that can be found for example in \cite{DS}. Again, the two 
$A_2^{(2)}$-KdV equations are given by Miura transformations; the one of our
interest is again $u(x)=-\phi'(x)^2 -\phi''(x)$. As shown before, central role
is played by the transfer matrix, wich is a solution of the associated linear
 problem $(\dx  - A_x(x;\la))T(x;\la)=0$. The formal solution is in this case 
\be T_{reg}(x,\la) =e^{h\phi(x)}{\cal P}
\exp\lt \int_0^xdy
(e^{-2\phi(y)} e_0+ e^{\phi(y)} e_1 ) \rt  .
\label{trama2}
\ee
The equation (\ref{trama2}) defines $T$ as an entire function of $\la$ with an essential
singularity at $\la=\infty$. The corresponding proper dressing symmetries may be worked 
out in a way similar to the $A_1^{(1)}$ case, but we are mainly
interested in the spectrum of local fields. As for the $A_1^{(1)}$ case, the
asymptotic expansion  is easily  written by following the general procedure
\cite{DS}. The result is: \be 
T_{asy}(x;\la)  = \left(\begin{array}{ccc} 1 & h^+_1 &  h^+_2 \\
                               0 &  1+h^0_3 &  h^0_1 \\
                               0 &  h^-_2 &  1+h^-_3 \end{array}\right)
\exp \lt -\int_0^x\sum_{i=0}^\infty f_i \Lambda^{-i} \rt  .
\ee
where $h^{\pm}_i,h^0_i$ are certain polinomials in $\phi'$, $f_{6k},f_{6k+2}$ are the 
densities of the local conserved charges of the $A_2^{(2)}$-KdV and $\Lambda=e_0+e_1$. 
Complying with our approach let us introduce the asymptotic resolvents 
\ba
Z_1=T_{asy}\Lambda T_{asy}^{-1} ; \nonumber \\
Z_2=T_{asy}\Lambda^2 T_{asy}^{-1}
\label{res2}
\ea
satisfying as before the equations (\ref{resdef}) $[\dx - A_x ,Z_i(x;\la)]=0$,
$i=1,2$ with  $A_x$ now given by (\ref{a22lax}) and (\ref{gen2}). These have
the form: \ba
Z_1 = \left(\begin{array}{ccc}     \la^{-1}a_1^{(1)}+\la^{-4}a_4^{(1)}+ \dots & 
      \la^{-2}b_2^{(1)}+\la^{-5}b_5^{(1)}+\dots &  1+\la^{-6}b_6^{(1)}+\dots \\
    1+\la^{-3}c_3^{(1)}+ \dots  &  -2\la^{-4}a_4^{(1)}+ \dots &
\la^{-2}b_2^{(1)}-\la^{-5}b_5^{(1)}+\dots  \\                                 
    \la^{-2}c_2^{(1)}+\dots    &  1-\la^{-3}c_3^{(1)}+\la^{-6}c_6^{(1)}+\dots 
& -\la^{-1}a_1^{(1)}+\la^{-4}a_4^{ (1)}+\dots  \end{array}\right) ,  \nonumber
\ea 
and 
\ba 
Z_2 = \left(\begin{array}{ccc}    
\la^{-2}a_2^{(2)}+\la^{-5}a_5^{(2)}+\dots  &
1+\la^{-3}b_3^{(2)}+\la^{-6}b_6^{(2)}+\dots &   \la^{-4}b_4^{(2)}+\dots \\ 
\la^{-1}c_1^{(2)}+\la^{-4}c_4^{(2)} \dots  &  -2\la^{-2}a_2^{(2)}+ \dots &
1-\la^{-3}b_3^{(2)}-\la^{-6}b_6^{(2)} +\dots  \\   1+\la^{-6}c_6^{(2)}+\dots  
 &  -\la^{-1}c_1^{(2)}+\la^{-4}c_4^{(2)}+\dots  &
\la^{-2}a_2^{(2)}-\la^{-5}a_5^{(2)}+\dots  \end{array}\right) \nonumber . \ea
For example, some expressions for the fiels in the entries of $Z_i, i=1,2$, as
a function of $v=-\phi'$ and its derivative, are: 
\ba 
a_1^{(1)} &=& -v ;
\nonumber \\ b_2^{(1)} &=& - \frac{1}{3} v^2 + \frac{1}{3} v' \virg 
c_2^{(1)}=- \frac{1}{3} v^2 - \frac{2}{3} v' ; \nonumber \\ c_3^{(1)} &=&
\frac{1}{3} v^3 + \frac{1}{3} vv' - \frac{1}{3} v'' \virg 
b_3^{(2)}=-\frac{2}{3} vv' + \frac{1}{3} v'' ; \nonumber \\ b_4^{(2)} &=&
-\frac{1}{9} v^4 + \frac{1}{9} v'^2 - \frac{2}{9} vv''+ \frac{2}{9} v'v^2 -
\frac{1}{9} v''', etc.  . 
\ea 
The equation (\ref{mKdV2}) is invariant under a
gauge transformation of the form (\ref{mkdvf}) . This latter will be a true
symmetry  provided the variation is proportional to $h$ : $\delta A_x= h
\delta \phi'$. We construct the  appropriate gauge parameters by means of the
resolvents (\ref{res2}) in a way similar to what we did in the
$A_1^{(1)}$-mKdV case: 
\be \theta_{6k+1}(x;\la)=(\la^{6k+1}Z_1(x;\la))_+ 
\virg   \theta_{6k-1}(x;\la)=(\la^{6k-1}Z_2(x;\la))_+ \label{theta's} \ee
which results in the following transformations for the  $A_2^{(2)}$-mKdV field
\be 
\delta_{6k+1}\phi'=\dx a_{6k+1}^{(1)} \virg   \delta_{6k-1}\phi'=\dx
a_{6k-1}^{(2)}  . \label{var2} 
\ee 
One can easily recognize in (\ref{var2})
the infinite tower of the commuting $A_2^{(2)}$-mKdV flows.

Now, in accordance with our geometrical conjecture, we would like to treat the entries of
the transfer matrix $T$ and of the resolvents $Z_i, i=1,2$ as independent fields 
and to build the spectrum of the local fields of $A_2^{(2)}$-KdV by means of 
them alone. As in the $A_1^{(1)}$ case, it turns out that not all of them are 
independent. If the defining relations of the resolvents are used, it is easy to see, that 
the entries of the lower triangle of both $Z_i$ can be expressed in terms of the rest.
Therefore, taking also into account the gauge symmetry of the system, one is led 
to the following proposal about the construction of the Verma module of the identity: 
\be
{\cV}^{mKdV}_{{\bf 0}}=\{l.c.o. \spz \delta_{6k_1+1}\dots 
\delta_{6k_M+1}\delta_{6l_1-1}\dots  
\delta_{6l_N-1}{\cP}(b_i^{(1)},b_j^{(2)},a_k^{(1)},a_l^{(2)})\},  
\label{veridm2} 
\ee
where $l.c.o.$ means {\it linear combinations of}.
Again, null-vectors appear in the r.h.s. of the (\ref{veridm2}) due to the constraints:
\be
Z_1^2=Z_2   \virg  Z_1Z_2=\buno
\label{con2}
\ee
and the equations of motion
\be
\delta_{6k\pm 1}Z_i=[\theta_{6k\pm 1},Z_i] \virg i=1,2  .
\label{eqsmot2}
\ee 
One can further realize that just as in the $A_1^{(1)}$ case, there is a subalgebra
consisting of the upper triangular entries of $Z_i, i=1,2$, closed under the action of 
the gauge transformations  $\delta_{6k\pm 1}$. The constraints (\ref{con2}) 
and (\ref{eqsmot2}) are consistent with such reduction giving a closed subalgebra of null vectors.
The first non-trivial examples are :
\ba
level 3 &:&  b_3^{(2)}-\dx b_2^{(1)}=0 ; \nonumber \\
level 4 &:&  b_4^{(2)}- (b_2^{(1)})^2 + \frac{2}{3} \dx b_3^{(2)} =0 ; \nonumber \\
level 6 &:&  b_6^{(1)} - 2 b_6^{(2)} + 2 b_2^{(1)}b_4^{(2)} + (b_3^{(2)})^2 =0 ; \nonumber \\
          &2&b_6^{(1)} -  b_6^{(2)} + 2 \dx b_5^{(1)} + \frac{1}{2} \dx^2 b_4^{(2)} + 
                           b_2^{(1)}b_4^{(2)} - (b_2^{(1)})^3 +(b_3^{(2)})^2 =0 .  
\ea
Therefore, in order to obtain the true spectrum of the family of the identity
(i.e. the $A_2^{(2)}$-KdV spectrum), one has to factor out, from the 
linearly generated Verma module
\be
{\cV}^{KdV}_{{\bf 0}}=\{l.c.o. \spz \delta_{6k_1+1}\dots
\delta_{6k_M+1}\delta_{6l_1-1}\dots \delta_{6l_N-1}
{\cP}(b_i^{(1)},b_j^{(2)})\} ,   \label{veridk2}
\ee
the Verma module of null-vectors ${\cN}^{KdV}_{{\bf 0}}$, i.e. 
\be
[{\bf 0}]={\cV}^{KdV}_0 / {\cN}^{KdV}_{{\bf 0}} .
\ee
Let us turn to the classical limit of the primary fields $e^{m\phi}, m=0,1,2,3,\dots$  . 
By virtue of the above reasoning  we conjecture for their Verma modules the expression
\be
{\cV}^{mKdV}_{{\bf 0}}=\{l.c.o. \spz \delta_{6k_1+1}\dots
\delta_{6k_M+1}\delta_{6l_1-1}\dots \delta_{6l_N-1}
[{\cP}(b_i^{(1)},b_j^{(2)})e^{m\phi}]\} .   \label{vermk2}
\ee
Again, we have to add to the null-vectors coming from (\ref{con2}) and (\ref{eqsmot2}) 
the new ones coming
from (repeated) application  of the equations of motion of the power $T^m(x;\la)$ 
\be
\delta_{6k\pm 1}T^m=\sum_{j=1}^mT^j\theta_{6k\pm 1}T^{m-j}
\label{eqmT^m2}
\ee 
obtaining the whole set of null-vectors ${\cN}^{KdV}_{{\bf m}}$. 
The first non-trivial examples of these additional null-vectors are given by:
\ba
level &2&: \spz (\dx^2 + 3b_2^{(1)}) e^{\phi} =0  \nonumber \\
level &3&: \spz (\dx^3 - 6 b_3^{(2)} + 12 \dx b_2^{(1)}) e^{2\phi} =0  \nonumber \\
level &4&: \spz (\dx^4 + \frac{135}{2} (b_2^{(1)})^2 + 30 \dx^2 b_2^{(1)} - \frac{27}{2} 
b_4^{(2)} - 30 \dx b_3^{(2)} ) e^{3\phi} =0 
\ea
where the operator $\dx$ acts on all the fields to its right. 
As a result, the ({\it conformal}) family of the primary field  $e^{m\phi}, m=0,1,2,3,\dots$ 
is conjectured to be in this case
\be
[{\bf m}]={\cV}^{KdV}_m / {\cN}^{KdV}_{{\bf m}}   .
\label{conmk2}
\ee

\resection{Conclusions and perspectives: off-critical theories and
quantisation.}

We have derived different types of symmetries of classical integrable systems
within a general framework. The unifying geometrical idea bases on dressing
simple objects by means of the transfer matrix $T$ with the aim to get the 
resolvent $Z$. In particular, the regular transfer matrix gives rise to
a Poisson--Lie symmetry of non-commuting flows. The construction of the 
spectrum employing this symmetry is an interesting open problem, clearly
connected with the spinon basis defined in \cite{I}, since the current density
$e^{\phi}$ is exactly the  spinon of \cite{I} after quantisation. Moreover,
it is of great interest the new way to look at the Sine-Gordon light-cone
evolution as generated by the sum of the first two regular vector fields. It
is also important to note that a similar half Virasoro symmetry can be obtained just 
underchanging the r\^ole of $x_-$ and $x_+$. We leave for a work in progress
the very important question on the whole algebra obtained from the union of
both half Virasoro algebras \cite{35}.

On the contrary, the asymptotic formula for the transfer matrix provides the
integrable hierarchy of (generalized) KdV flows and, in the case of dressing of
non-cartan generators, two new infinite series of flows closing an algebraic
structure with the integrable ones. A new costruction of the spectrum of
classical Virasoro algebra has been given in terms of these ingredients.

Even in view of quantisation, it is worth extending our constructions out of criticality 
in another way. We may define the anti-chiral transfer matrix
\be
\bar T(\bar x,\la) ={\cal P}
\exp\lt \la \int_x^0dy
(e^{-2{\bar \phi}(y)} E+ e^{2{\bar \phi}(y)} F ) \rt e^{H{\bar \phi}(\bar x)}
\label{atm}
\ee
which solves the linear problem
\be
\p_{\bar x} \bar T(\bar x;\la) =\bar T(\bar x;\la)\bar A_{\bar x}(\bar x;\la)
\ee
where $\bar A_x(x;\la)$ is obtained from $A_x$ by substitution $\phi\rightarrow -\bar\phi$. 
In accordance with \cite{BLZ}, we suggest for the off-critical transfer matrix
\be
{\bf T}(x,\bar{x};\la;\mu) = \bar{T}(\bar{x};\mu/\la)T(x;\la),
\ee 
and correspondingly for ${\bf Z}(x,\bar x; \mu ;\lambda)$.

Following the basic work \cite{BLZ}, one  quantizes 
the corresponding mKdV system by replacing the Kac-Moody algebra with the corresponding 
quantum group and the mKdV field $\phi$ with the
Feigin--Fuchs--Dotsenko--Fateev free field \cite{FF}. As explained in
\cite{FRS} the importance of  considering also the $A_2^{(2)}$-mKdV
hierarchies is due to the fact that the  quantisation of this second
semiclassical system exhausts the integrability  directions of theories of
type  (\ref{pert}) starting from Minimal Models of \cite{BPZ}.  For different
kinds of CFT it is sufficient to considere hierarchies attached to different
Kac-Moody algebras (in the Drinfeld-Sokolov scheme \cite{DS}).

In conclusion, we have presented a generalization of the dressing symmetry
construction  leading to a 
non-local Virasoro symmetry of the mKdV hierarchy and SGM. We stress that it
has nothing to do with the space-time Virasoro, generated at the classical
level by the moments of the {\it classical stress tensor} $\int x^n u(x)dx$.
It is obtained instead by dressing the differential operator
$\lambda^{n+1}\partial_{\lambda}$. In view of the relation between the
spectral parameter and the on-shell rapidity $\lambda= e^ \theta$, it is
generated probably by diffeomorphisms in the momentum space and in this sense
is dual to the space-time Virasoro symmetry. Although we presented a
construction only in the case of mKdV, it can be easily extended for the
generalized KdV theories as well. Of particular interest is the $A^{(2)}_2$
hierarchy, connected with the $\phi_{1,2}$ perturbation of CFT models
\cite{FRS}.

Furthermore, such a symmetry appears also in the study of Calogero-Sutherland model 
whose connection 
with the matrix models and CFT is well known. Moreover, it is known that in
the q-deformed case it becomes a deformed Virasoro algebra \cite{SKAO}. It is 
natural to suppose that in the same way our construction is deformed
off-critically.

We suggest that this Virasoro symmetry could be of great importance for the 
study of {\bf 2D-IQFT}. First of all it should provide a new set of conserved
charges closing a  non-abelian algebra, 
thus carrying nessesarily more information about the theory. Furthermore one
might quantise these charges at conformal and off-critical level. We
have reasons to belive that the perturbed version (i.e. SGM) should be closely
related to the aforementioned DVA. 

Recently, Babelon, Bernard and Smirnov \cite{BBS} constructed certain null-vectors off-criticality 
in the context of the form-factor approach. They showed that there is a deep
connection, at the classical level, between their construction and the finite
zone solutions of KdV and the Witham theory of averaged KdV. On the other
hand the Virasoro algebra presented above has a natural action on the finite
zone solutions, changing the complex structures of the corresponding
hyperelliptic Riemann surfaces, and  on the basic objects of the Witham
hierarchy \cite{GO}. This suggests we may found the quantum action of our symmetries 
(in particular the Virasoro one) in SGM using the form factor formalism
developed in \cite{BBS}.

\vspace{1cm}

{\bf Acknowledgments} - We are indebted to E. Corrigan, P. Dorey, G.
Mussardo, I. Sachs and F. Smirnov for discussions and interest in this work.
D.F. thanks the I.N.F.N.--S.I.S.S.A., the Mathematical Sciences Departement
in Durham and the EC Commission (TMR Contract ERBFMRXCT960012) for financial
support. M.S. acknowledges S.I.S.S.A. for the warm hospitality over part of
this work.

\end{document}